\documentclass[12pt]{article}

\usepackage{amsmath,amssymb,amssymb,graphicx}
\usepackage{amsbsy}
\usepackage{cite}
\usepackage{hyperref}

\newcommand{\arXiv}[2]{\href{http://arxiv.org/pdf/#1}{{\tt #2/#1}}}
\newcommand{\arXivold}[1]{\href{http://arxiv.org/pdf/#1}{{\tt #1}}}
\newcommand{\beq}{\begin{eqnarray}}
\newcommand{\eeq}{\end{eqnarray}}

\newcommand{\centeron}[2]{{\setbox0=\hbox{#1}\setbox1=\hbox{#2}\ifdim
\wd1>\wd0\kern.5\wd1\kern-.5\wd0\fi \copy0
\kern-.5\wd0\kern-.5\wd1\copy1\ifdim\wd0>\wd1
                                   \kern.5\wd0\kern-.5\wd1\fi}}
\newcommand{\ltap}{\>\centeron{\raise.35ex\hbox{$<$}}
                           {\lower.65ex\hbox{$\sim$}}\>}
\newcommand{\gtap}{\>\centeron{\raise.35ex\hbox{$>$}}
                           {\lower.65ex\hbox{$\sim$}}\>}

\newcommand\ZZ{\hbox{\zfont Z\kern-.4emZ}}
\font\zfont = cmss10 

\newcommand{\drawsquare}[2]{\hbox{%
\rule{#2pt}{#1pt}\hskip-#2pt
\rule{#1pt}{#2pt}\hskip-#1pt
\rule[#1pt]{#1pt}{#2pt}}\rule[#1pt]{#2pt}{#2pt}\hskip-#2pt
\rule{#2pt}{#1pt}}

\newcommand{\fund}{\drawsquare{6.5}{0.4}}

\newcommand{\asymm}{\raisebox{-3.5pt}{\drawsquare{6.5}{0.4}\hskip-6.9pt%
        \raisebox{6.5pt}{\drawsquare{6.5}{0.4}}}}

\begin{document}
\begin{titlepage}

\vskip.5cm \begin{center} {\huge \bf Light Stops from \
\vskip 8pt
Seiberg Duality}\\ \vskip
8pt

\vskip.1cm \end{center} \vskip0.2cm

\begin{center} {\bf {Csaba Cs\'aki}$^{a}$, {Lisa Randall,}$^{b}$ {\rm
and} {John Terning}$^{c}$} \end{center} \vskip 8pt

\begin{center} $^{a}$ {\it  Department of Physics, LEPP, Cornell
University, Ithaca, NY 14853} \\ \vspace*{0.1cm} $^{b}$ {\it Department
of Physics, Harvard University, Cambridge, MA 02138} \\ \vspace*{0.1cm}
$^{c}$ {\it Department of Physics, University of California, Davis, CA
95616} \\

\vspace*{0.3cm} {\tt  csaki@cornell.edu, randall@physics.harvard.edu, \\
terning@physics.ucdavis.edu} \end{center}

\vglue 0.3truecm

\begin{abstract}
\vskip 3pt
\noindent If low-energy supersymmetry is
realized in nature,  a seemingly contrived hierarchy in the squark mass
spectrum appears to be required. We show that composite supersymmetric
 theories at the bottom of the conformal window can automatically yield the spectrum that is suggested by
experimental data and naturalness. With a non-tuned choice of parameters,  the only
superpartners below one TeV will be the partners of the Higgs, the
electroweak gauge bosons, the left-handed top and bottom, and the
right-handed top, which are precisely the particles  needed to make weak
scale supersymmetry breaking  natural. In the model considered here, these correspond to
composite (or partially composite) degrees of freedom via Seiberg
duality, while the  other MSSM fields, with their heavier superpartners,
are elementary. The key observation is that at or near the edge of the conformal
window, soft supersymmetry breaking scalar and gaugino masses are
transmitted only to fundamental particles at leading order. With the
potential that arises from the duality,  a Higgs with a 125 GeV mass,
with nearly SM production rates, is naturally accommodated without  tuning. The lightest ordinary superpartner is either the
lightest stop or the lightest neutralino. If it is the stop,  it is
natural for it to be almost degenerate with the top, in which case it
decays to top by emitting a very soft gravitino, making it quite
difficult to find this mode at the LHC and more challenging
to find SUSY in general, yielding a simple
realization of the stealth supersymmetry idea. We analyze four benchmark
spectra in detail.
\end{abstract}

\end{titlepage}


\section{Introduction}
\label{sec:intro}
\setcounter{equation}{0}
\setcounter{footnote}{0}

Supersymmetry potentially provides a complete theory of electroweak symmetry
breaking, eliminating the hierarchy problem for the Higgs mass. But in a way
supersymmetry is too efficient in suppressing the Higgs mass: the
natural mass for a SUSY Higgs is often below 100 GeV so that large
radiative corrections become essential. The simplest versions of the
supersymmetric extension of the Standard Model (SM) are now being
severely challenged: the Higgs sector must be fine tuned at the
sub-percent level in order to push the Higgs mass sufficiently far above
the $Z$ mass, and the non-observation of missing energy events at the
LHC~\cite{CMSSUSY,ATLASSUSY} puts impressive  bounds on squark and
gluino masses. In popular versions of the Minimal Supersymmetric
Standard Model (MSSM) with degenerate squarks, these masses are now
constrained to be above 1 TeV. Minimizing fine-tuning in light
of this data requires that the stop  squark is  lighter than
the first and second generation
squarks~\cite{lightstopGMSB,lightstopRaman,lightstopAndi,lightstopSLAC}, leading to
yet another hierarchy within SUSY models. The aim of this paper is to
present a model where both the squark mass hierarchy and the little
hierarchy are solved naturally via compositeness.

Compositeness is an intriguing idea for electroweak symmetry breaking:
strong dynamics could either directly break electroweak symmetry or
produce a composite Higgs boson without a hierarchy problem.  Flavor poses the biggest challenges for such models, but
compositeness might actually explain the much greater mass of the the top quark:  if the $t$ and Higgs are
composite while other quarks are not,  then their Yukawa coupling is generically order one, while the
other Yukawa couplings must be generated by higher dimensions operators.
A fully composite SM
(like that proposed by Abbott and Farhi \cite{Abbott})  is not expected
to yield weakly interacting $W$s and $Z$s. However we have learned
from warped extra dimensional models  (which may be duals of
approximately conformal 4D theories), like the Randall-Sundrum (RS) model
\cite{RS},  that large anomalous dimensions can save the composite Higgs
scenario at the price of having both an elementary and a composite
sector present, and having the $t$ quark only partially composite (along
with the $W$ and $Z$). Even for these models,   some fine tuning is nonetheless required
to make the composite Higgs much lighter than the composite $W^\prime$.

Since the problems of SUSY and of compositeness are complementary, it
seems natural to try to combine the two to produce one complete, natural
model of EWSB at the TeV scale. In general this might seem artifical but
existing Seiberg dualities automatically feature both.  We will see in the models we consider that
not only do we get the best features of both models, but also that supersymmetry breaking decouples at leading
order from the IR composite states, somewhat analogously to what happens with UV supersymmetry breaking in RS-type models,
leading to a natural hierarchy in the superpartner spectrum that readily accommodates current constraints.

Other ideas that
have been explored include refs. \cite{fatHiggs1,fatHiggs2,MCSSM,similar},  in
which strong SUSY dynamics trigger electroweak symmetric breaking by
producing a composite Higgs that obtains a VEV. Generically if the model
reduces to the MSSM when the strong SUSY scale is taken to be much
larger than the electroweak scale then the problems of the MSSM are
reproduced.  If, however, the model reduces to the Next-to-Minimal
Supersymmetric Standard Model (NMSSM) when the strong SUSY scale is
taken to be large, then the Higgs mass can be much larger
\cite{fatHiggs1,fatHiggs2,MCSSM,similar}  than in the MSSM or even the
NMSSM.  This is because the cubic coupling between the composite singlet
and composite Higgs doublets, which tends to increase the Higgs mass,
can be much larger than in the NMSSM since its Landau pole merely
signals the existence of the strong SUSY dynamics that generated the
cubic coupling of the composites in the first place. This will be true for our
 model as well (as in~\cite{MCSSM}) and allows for sufficiently  heavy Higgses.

Following the SUSY compositeness idea further, one must account for
SUSY breaking.
In this paper we show that
(perhaps unexpectedly) the composite superpartners can remain light
while the elementary superpartners can be heavy. One elegant idea for
addressing SUSY breaking is to have strong SUSY dynamics break SUSY
as well \cite{SingleSector,similar}. Such models are referred to as
single sector models. In this case however, the composites of the strong SUSY
sector have large SUSY breaking masses. Avoiding fine-tuning calls
for a light stop, ${\tilde t}$, so in these single sector models the $t$
 quark must be elementary, and thus the Higgs should be elementary as
well in order to get a large $t$ quark mass.  Thus if we want a
relatively light composite ${\tilde t}$ as well as  a composite $t$ and
Higgs, SUSY breaking must come from outside the strong
sector that produces composites. In this case, the leading
contributions to the composite soft masses are calculable
\cite{Cheng,ArkaniHamed:1998kj,NimaRiccardo,MarkusRiccardo,Abel} when the
Seiberg dual is weakly coupled in the infrared.

Generically the results
are discouraging \cite{Cheng} since the squared soft masses of the
mesons and dual quarks add to zero, so at least some of the composites
will be tachyonic. However, at the boundary of the conformal window, the
leading contributions to soft masses vanish.  This suggests an
interesting hierarchy of soft breaking masses:  the composites (like the
Higgsinos, ${\tilde t}_L$, ${\tilde b}_L$, and ${\tilde t}_R$) are much
lighter than the other superpartners.  If the $W$ and $Z$ are also
partially composite, then their superpartners can also be lighter than
the elementary superpartners. Note that these are exactly the particles
that are needed to cancel the quadratic divergence in the Higgs mass. In
this case the (approximately conformal) strong dynamics shields the
composites from large supersymmetry breaking.  On the other hand the
little hierarchy problem of composite Higgses is resolved here via
supersymmetry: the Higgs is a dual quark of Seiberg duality that
can be much lighter than the compositeness scale because of supersymmetry. Moreover,  because
of the form of the potential that arises from Seiberg duality, the typical mass is of  order the
Higgs VEV without the usual MSSM suppression by a gauge coupling.  Therefore in this model it is possible to accommodate a Higgs mass of 125 GeV without any tuning, while the production and decay rates of the Higgs will be close to SM values. In fact, the recently presented hints for a 125 GeV Higgs from ATLAS~\cite{ATLASHiggs} and CMS~\cite{CMSHiggs} might even be further evidence that a viable supersymmetric theory should incorporate a low-scale cutoff, such as the compositeness scale presented here. 
The resulting spectrum is
reminscent of the ``more minimal supersymmetric standard model" idea
of~\cite{Cohen}. It can also be viewed as an explicit four-dimensional
implementation of the warped extra dimensional supersymmetric models
of~\cite{Tony1,Raman,Tony2}.

In this paper we analyze such models, which  arise as dual composite gauge
theories at the edge of the conformal window. These models have
(partially) composite Higgs, $t$, $W$ and $Z$ and can address three
problems at once: the hierarchy of Yukawa couplings,  the little
hierarchy of the Higgs mass, and the hierarchy of the squark soft
masses.  The same composite
states that are needed for a dynamical Yukawa coupling are the ones
needed to protect the Higgs mass.
Such an unconventional superpartner
spectrum has important consequences for SUSY searches at the LHC.

 A
limiting case is a nearly degenerate $t$-${\tilde t}$ sector \cite{Kats} that could be
 naturally produced by compositeness.  In this case all superpartners
decay via the NLSP ${\tilde t}$ and not much missing energy. Models with
new approximately degenerate superpartners that end the decay chains of
Standard Model superpartners have been termed stealth SUSY models
\cite{Fan} precisely because of this lack of missing energy signatures.
In generic  stealth SUSY models, the approximate degeneracy is caused by
a suppression of the coupling of the new states to the SUSY breaking
sector.  In the composite models we are discussing here, the suppression
arises precisely because the states are composites of the strong SUSY
dynamics, and the almost conformal strong dynamics screens
SUSY breaking:  the anomalous dimensions of the supersymmetry breaking terms
suppress them  up to possible threshold corrections.  The threshold
corrections are determined by holomorphy and also vanish in the conformal window.

Interestingly, the recently proposed \cite{MCSSM} Minimal
Composite Supersymmetric Standard Model (MCSSM)  has just these
composite degrees of freedom and sits on the edge of the conformal
window, so it provides a benchmark model for exploring this scenario
and we focus on that model in this paper.
Through most of the paper we will assume a low-scale mediation scenario,
for which the prime example is gauge mediation. Many of the
problems of gauge mediation simply do not arise here, since we can break
electroweak symmetry in the SUSY limit there is no $B \mu$
problem, while singlet
soft breaking terms are easily obtained since the singlet is a
composite. We also consider one example of a possible high scale supersymmetry breaking model as well.

Some of the key ideas here can be understood in analogy to   the RS picture, where composites are localized near an IR brane, while the elementary fields are on the UV brane. The insensitivity of composites to SUSY breaking is simply captured by a small overlap of the IR localized composites with the UV localized SUSY breaking~\cite{Tony1}. The other main ingredient is partial compositeness, which solves the major problems of fully composite theories. This is another very familiar feature of realistic RS models,  corresponding to (almost) flat wave functions~\cite{bulkgauge} for the $W$ and $Z$.

The paper is organized as follows. First we discuss how external SUSY
breaking feeds through to the composites of Seiberg duality. In
section~\ref{sec:mcssm} we review the MCSSM~\cite{MCSSM}, which is the simplest
model with composite Higgses, $t$'s and partially composite $W$s and $Z$s. In
section~\ref{sec:parameters} we estimate the sizes of realistic
parameters for the MCSSM, discuss the electroweak symmetry breaking
potential, and present the mass matrices for the light sparticles. In
section~\ref{sec:stealth} we discuss the phenomenology by focusing
on four benchmark spectra. Two of them have ${\tilde t}$ NLSP's, and one of these two has the lightest ${\tilde t}$ almost
degenerate with the $t$ as in~\cite{Fan} and therefore can be
kinematically accessible to the current LHC run while nonetheless
avoiding detection so far, while the other has a somewhat heavier ${\tilde t}$
below 300 GeV.  The other two spectra are more conventional with neutralino NLSP's,
one of which corresponds to a gauge mediated spectrum. All four of the spectra have $\tan\beta \sim 1$,
and $\tan \beta$ can even be smaller than 1.

\section{Soft Breaking Terms for Composites}
\label{sec:soft}
\setcounter{equation}{0}
\setcounter{footnote}{0}

Before we present the concrete composite SUSY model that solves both the
little hierarchy problem and predicts light ${\tilde t}$'s, we first address the
question of the magnitudes of the soft breaking terms in composite SUSY
models. Since these will essentially determine the characteristics of
the spectrum, this is the critical feature of this class of models. We will
assume that the strong dynamics can be captured via Seiberg duality,
and ask the question of how UV soft breaking terms for the
elementary (``electric") degrees of freedom get transmitted to the
composite (``magnetic") degrees of freedom.    We  apply the
method of analytic continuation into
superspace~\cite{ArkaniHamed:1998kj,NimaRiccardo,MarkusRiccardo} to find the mapping
of soft breaking terms under duality.  We start with the Lagrangian for
the electric quarks ${\cal Q,\bar{Q}}$ of an electric $SU(N)$ gauge
theory with $F$ flavors of these quarks.

We want to compare the soft mass for some spectator  ``elementary" degrees of freedom  that do not have strong interactions with the soft
masses of the composites in the IR. From the RS picture, we expect that composites (localized in the IR)
will be insensitive to soft SUSY breaking in the UV,  while the elementary fields should be sensitive. Indeed the soft  breaking masses for the elementary fields undergo a perturbative RG running between the UV and IR scales characterized by small perturbative anomalous dimensions
\beq
m_{el}^2(\mu) = m_{UV}^2 \left(\frac{\mu}{\Lambda}\right)^{{\mathcal O}(\alpha)}~,
\eeq
up to perturbative threshold corrections.
The composite fields can in principle  have both non-perturbative finite terms and non-perturbative  anomalous dimensions
\beq
m_{comp}^2(\mu) = m_{IR}^2 +m_{UV}^2\left(\frac{\mu}{\Lambda}\right)^{\gamma}~,
\label{non-pertRG}
\eeq
This equation is schematic, when $\gamma$ is a function of $\mu$, then the RGE solution has the form of an exponential of an integral of $\gamma(\mu)$. Unlike the running term, the interpretation of the finite threshold term $m^2_{IR}$ is not immediately obvious in the RS picture. When the dual theory is weakly coupled in the IR, the finite term, $m_{IR}^2$, can be calculated
\cite{Cheng,NimaRiccardo} using holomorphy. Meanwhile, the existence of a well-behaved Seiberg dual requires that the anomalous dimension, $\gamma(\mu)$, is  positive and for a weakly coupled dual,
$\gamma(\mu) \sim {\mathcal O}(1)$ in the IR.  A large positive anomalous dimension rapidly drives the second term in (\ref{non-pertRG}) to zero.
This analysis extends into
the conformal window as well~\cite{MarkusRiccardo}, where it further can be shown \cite{Abel} that $m_{IR}=0$.  This simply means that the fixed point is attractive, and these soft mass terms are irrelevant and vanish at the fixed point. As a consequence, one expects $m_{IR}$ to also vanish just at the boundary between the  conformal window and the free magnetic phase.  At the bottom end of the conformal window $\gamma(\mu)$ is still ${\mathcal O}(1)$ but is perturbative at the top of the window, which means that at the top of the window the approach to the fixed point can be very slow,  and in this case one exits the RGE long before the fixed point is approached.  In the free magnetic phase similar conclusions hold, but with  $m_{IR}\ne0$ in general,  as one can see from the low-energy effective K\"ahler potential \cite{NimaRiccardo}. As we shall see, the two approaches agree at the bottom edge of the conformal window.

Next we will explicitly show the calculation for $m^2_{IR}$ in the weakly coupled, ``free-magnetic" phase.  We will also include a small
supersymmetric mass (matrix)  $\mu_f$ for the electric quarks, much
smaller than the dynamical scale of the theory. One of these will correspond to the term triggering electroweak symmetry breaking, which in this model will happen even in the absence of supersymmetry breaking, but via the composite dynamics. Thus one needs to assume that the relevant $\mu_f$ is related to the magnitude of the Higgs VEV $v$, and this parameter is what sets the electroweak scale. Although we do not explain this choice of parameter, we expect that in   a more complete model of supersymmetry breaking this can be related to the soft supersymmetry breaking scale as well.

The effects of the soft SUSY
breaking terms for the elementary fields are incorporated into the
Lagrangian by using the real and chiral spurions $Z$ and $U$ with
non-zero $\theta$ components~\cite{Cheng,ArkaniHamed:1998kj,NimaRiccardo,MarkusRiccardo}:
\beq
{\mathcal L} = \int d^4 \theta \left( {\mathcal Q}^\dagger Z e^V
{\mathcal Q}+{\bar {\mathcal Q}}^\dagger Z e^V {\bar {\mathcal
Q}}\right) + \int d^2\theta \left( U W^\alpha W_\alpha + \mu_f {\bar
{\mathcal Q}}{\mathcal Q}\right) +h.c. \ .
\eeq

To introduce a soft squark mass $m_{UV}$, a gaugino mass $m_\lambda$, and a soft-breaking $B$ term (with $m_{UV}^2\sim m_\lambda^2\sim B$)  we Taylor
expand the spurions in superspace coordinates:
\beq
Z&=& 1-\theta^2 B-{\bar \theta}^2 B- \theta^2 \bar\theta^2 (m_{UV}^2-|B|^2)\\
U&=& \frac{1}{2
g^2}-i \frac{\theta_{YM}}{16 \pi^2}+\theta^2 \frac{m_\lambda}{g^2}~,
\eeq
where we have also included the CP violating parameter $\theta_{YM}$ (not to be confused with the superspace coordinate).
The spurion $U$ is related to the holomorphic strong
scale $\Lambda_h$ which acts as a chiral superfield spurion that is also
an RG invariant:
\beq
\Lambda_h=\mu \,e^{- 16 \pi^2 U(\mu)/b}
\eeq
where
$b$ is the one-loop $\beta$-function coefficient $b= 3N-F$ and $\mu$ is the RG scale.
In the model presented in the next section we will choose $N=4$
and $F=6$.

We can also include these spurions in the composite description since
the structure of the low-energy theory is constrained by symmetries
including an anomalous axial $U(1)$ symmetry. In other words $Z$ and
$U$ are also spurions of the anomalous axial $U(1)$. Under axial
transformations, where the rotation parameter is promoted to a chiral
superfield $A$, we have
\beq
&&{\mathcal Q}\rightarrow e^{A}{\mathcal
Q}~, \quad \quad {\bar {\mathcal Q}} \rightarrow e^{A}{\bar {\mathcal
Q}} \\ &&Z \rightarrow e^{-A-A^\dagger}~, \quad \quad \Lambda_h
\rightarrow e^{2 F/b\, A} \Lambda_h
\eeq
It is convenient to  introduce
a redundant scale that is invariant under axial transformations
\beq
\Lambda^2 = \Lambda_h^\dagger Z^{2F/b} \Lambda_h
\eeq
which is also a
SUSY breaking spurion
\beq
\log \frac{\Lambda}{\mu}= \frac{- 8 \pi^2 }{b
g^2} +\frac{- 8 \pi^2 m_\lambda}{b g^2} (\theta^2 +{\bar \theta}^2) -
\frac{F}{b} m_{UV}^2 \theta^2 {\bar \theta}^2\ .
\eeq
This $\Lambda$ is the invariant scale that can be used
for dimensional analysis once the anomalous $U(1)$ charge is fixed.

In the composite theory, "magnetic" states transform under the dual gauge $SU(F-N)$
gauge group, and include the meson $M$ and dual quarks $q,\bar{q}$. Due
to the operator mapping
\beq
{\mathcal Q}{\bar {\mathcal Q}}
\leftrightarrow  M~,\quad\quad {\mathcal Q}^N  \leftrightarrow
q^{F-N}~,\quad\quad {\bar {\mathcal Q}}^N  \leftrightarrow  {\bar
q}^{F-N}
\eeq
we have the following axial transformations for the
composite states:
\beq
q \rightarrow e^{A N/(F-N)}q\\ {\bar q}
\rightarrow e^{A N/(F-N)}{\bar q}\\ M \rightarrow e^{2 A} M~.
\eeq
Since the dual composite theory is in the weakly coupled phase we can write an approximately canonical K\"ahler potential. Requiring SUSY  and axial invariance and using dimensional analysis we
find the dual Lagrangian
\beq
{\mathcal L} &=& \int d^4 \theta \left[
\frac{M^\dagger Z^2 M}{\Lambda^2} +\frac{q^\dagger Z^{N/(F-N)} e^{\tilde
V} q}{\Lambda^{(4N-2F)/(F-N)} } +\frac{{\bar q}^\dagger Z^{N/(F-N)}
e^{\tilde V} {\bar q}}{\Lambda^{(4N-2F)/(F-N)} } \right]  \nonumber \\
&&+ \int d^2\theta \left[ \,U {\widetilde W}^\alpha {\widetilde
W}_\alpha+\frac{y\,M q {\bar q}}{\Lambda_h^{b/(F-N)}}+\mu_f M \right]
+h.c.
\label{eq:dualLagrangian}
\eeq
We can read off the soft masses near the infrared fixed point
\cite{NimaRiccardo,MarkusRiccardo} for the composites from the K\"ahler
term by Taylor expanding in superspace:
\beq
m_M^2= 2\frac{3N-2F}{b}
m_{UV}^2~,\quad \quad  m_q^2 =-\frac{3N-2F}{b}m_{UV}^2 \label{soft}
\eeq
Generically these results spell trouble for composite models: some of
the dual quark or meson soft breaking masses should be tachyonic, and
this would apply for the entire multiplet. However, for the case when
$F=3N/2$, that is at the lower end of the conformal window these leading
calculable terms vanish. This is exactly the right region for the model
considered later in this paper ($F=4,N=6$).
In this case the leading terms will come from the second term in (\ref{non-pertRG})
corresponding to the fact that we do not run all the way to $\mu=0$ but stop at a scale  given by (\ref{non-pertRG}) $\mu^2 \sim m_{UV}^2 \mu/\Lambda$, so that the  corrections are ${\mathcal O}\left(m_{UV}^4/\Lambda^2\right)$  which can also be seen as the effects of higher order terms
in the K\"ahler potential suppressed by additional powers of $\Lambda$ \cite{NimaRiccardo}.
The perturbative dual gauge group corrections are included in this estimate. In addition to power corrections, there are also perturbative corrections from
SM gauge interactions that could dominate when $\Lambda$ is very large.

The matching of the gaugino masses follows simply from the invariance of
$\Lambda$, implying
$m_{\lambda}/(bg^2)=m_{\tilde{\lambda}}/(\tilde{b}\tilde{g}^2)$ in the
holomorphic basis. After the rescaling by couplings to get into the
canonical basis one obtains the well-known answer
\beq
m_{\tilde\lambda}= -\frac{3N-2F}{3N-F} m_\lambda~,
\label{gauginomatched}
\eeq
thus
the leading contribution of the composite gaugino mass also vanishes at
the boundary of the conformal window.

To get the soft terms that come from the superpotential couplings we
must rescale the fields to get canonical K\"ahler terms. Since we
need terms only of order $\theta^2$ we can write
\beq
Z= \xi^\dagger
\xi~,\quad\quad \xi = 1-\theta^2 B
\eeq
and then rescale chiral fields
only via the holomorphic quantities $\xi, \Lambda_h$. We then find the
superpotential terms in the canonical basis:
\beq
\int d^2\theta \left( \,y\,
M q {\bar q}  +  \mu_f \Lambda_h M \xi^{\frac{2(2F-3N)}{(3N-F)}} +h.c. \right)
\label{eq:superpotmatched}
\eeq
Since the cubic superpotential is independent of the supersymmetry breaking spurions, we find that the $A$-term vanishes in the IR limit for any $F$:
\beq
A= {\cal O}(\frac{m_{UV}^2}{\Lambda})\ .
\eeq
We also find a SUSY breaking scalar tadpole for the meson
\beq
T= \mu_f \Lambda
\left( -\frac{16 \pi^2 m_\lambda}{b g^2} -\frac{2(2F-3N)}{3N-F} B\right) \ .
\label{eq:tadpole}
\eeq
While the second term vanishes for $F=3/2N$ the first one does not: this is not surprising since this is the effect of an explicit breaking of the conformality on the elementary side.
The expected magnitude for $T$ will then be of order
\begin{equation}
T \sim \mu_f \Lambda \times m_{UV},
\end{equation}
where $m_{UV}$ represents the characteristic magnitude of the gaugino mass $m_\lambda$ that appears on the right hand side of equation (\ref{gauginomatched}).
 Thus we find that the IR limit of all soft breaking parameters for composites vanish at the edge of the conformal window, except for the scalar tadpole, which is related to the explicit breaking term and the elementary SUSY breaking terms.  For phenomenological reasons that will be explicit in the next section we parameterize the superpotential term linear in the meson field in (\ref{eq:superpotmatched}) as $y f^2 M$, where $f$ must be chosen to be of order of the weak scale, and $y$ is
the dynamical Yukawa coupling that runs down to ${\cal O}(1)$ at the electroweak scale, which is the right size to give the correct $t$ mass.
In terms of the duality mapping given above, we see that by definition
$y\,f^2\equiv  \mu_f \Lambda$, so we find that the magnitude of the scalar tadpole is of order
\beq
T \sim f^2 m_{UV}
\eeq
Thus we find that at the edge of the
conformal window one has a hierarchy of the soft breaking terms, which,
writing the soft scale for the elementary fields as  $m_{el}\sim m_{UV}$, takes
the form
\beq
 A, m_{\tilde{q},\tilde{g}} &\sim&
\frac{m_{el}^2}{\Lambda}\ll m_{el} \nonumber \\
T & \sim & \mu_f \Lambda \times m_{el}\equiv f^2 m_{el} \ll m_{el}^3~.
\eeq

As a check of the duality mapping, note that the scale matching relation
between the electric and dual magnetic theories is defined in the frame
where the dual quarks  are canonically normalized, and the meson is
mapped to ${\cal Q \bar{Q}}$. In this frame the dual quarks carry
anomalous charge 1, and the scale matching relation is~\cite{Seiberg}
\beq
\Lambda_h^b {\tilde
\Lambda}_h^{\tilde b}=(-1)^N \Lambda_{\tiny M}^F
\eeq
where $\Lambda_{\small M}$ can be
expressed in terms of $\Lambda_h$ and $\xi$ by matching the anomalous charge as:
\beq
\Lambda_{\small M} = \Lambda_h
\xi ^{\frac{3(2N-F)}{3N-F}}
\eeq
By rescaling the terms in (\ref{eq:dualLagrangian}) to move to a frame with canonically normalized dual quarks we find that as expected $\Lambda_{\small M}$
is also the parameter appearing in the dual superpotential in this
frame: $Mq\bar{q}/\Lambda_{\small M}$, as predicted in~\cite{Seiberg}.

\section{MCSSM: The Model for a Composite Third Generation}
\label{sec:mcssm}
\setcounter{equation}{0}
\setcounter{footnote}{0}

A concrete model (that we refer to as the Minimal Composite
Supersymmetric Standard Model or MCSSM) of supersymmetric composite
Higgs and $t$ quarks (and partially composite $W$ and $Z$)   was
recently proposed in \cite{MCSSM}. The main idea is that an
asymptotically free gauge group becomes strongly interacting and the IR
theory will contain composite gauge bosons, mesons and dual quarks, some
of which are to be identified with the $W$, $Z$, $t$, and Higgs of the MSSM.
To get a realistic theory, the composite $W$ and $Z$ need to be mixed
with elementary $W$ and $Z$ gauge bosons that couple to the elementary
quarks and leptons. The electric theory of the simplest such model is
given by (corresponding to $N=4,F=6$)
\beq
\begin{array}{c|c|cccc} &
SU(4) & SU(6)_1  & SU(6)_2 & U(1)_V & U(1)_R \\
\hline {\cal Q}& \fund &
\overline{\fund}   & {\bf 1} & 1  & {\frac{1}{3}}
\vphantom{\raisebox{3pt}{\asymm}}\\ {\cal \bar Q} & \overline{\fund}   &
{\bf 1}  & \overline{\fund} & -1 & {\frac{1}{3}}
\vphantom{\raisebox{3pt}{\asymm}}\\ \end{array}
\label{electricfinal}
\eeq
where the $SU(4)$ is the strong gauge group and the other groups
are the global symmetries, some of which are weakly gauged.  In particular, the elementary gauge symmetries $SU(3)\times SU(2)_{el}\times U(1)$ are embedded into these global symmetries. We will also allow small tree-level masses for the electric quarks.

The IR behavior of this strongly coupled theory is given by the Seiberg
dual~\cite{Seiberg}
\beq
\begin{array}{c|c|cccc} & SU(2)_{\rm mag} & SU(6)_1  & SU(6)_2
& U(1)_V & U(1)_R \\ \hline q& \fund & \fund & {\bf 1} & 2  &
{\frac{2}{3}} \vphantom{\raisebox{3pt}{\asymm}}\\ \bar q&
\overline{\fund}   & {\bf 1}  & \fund & - 2 & {\frac{2}{3}}
\vphantom{\raisebox{3pt}{\asymm}}\\ M& {\bf 1} &  \overline{\fund}   &
\overline{ \fund} & 0 & {\frac{2}{3}}
\vphantom{\raisebox{3pt}{\asymm}}\\ \end{array}
\label{magneticfinal}
\eeq
with the additional dynamical superpotential term
\beq
W_{dyn}= y
\, \bar q M q~. \label{dualsuper}
\eeq

The SM gauge groups are  embedded in the global symmetry as
\beq
\begin{split} &SU(6)_1 \supset SU(3)_{c} \times
SU(2)_{\rm el} \times U(1)_Y \\ &SU(6)_2 \supset SU(3)_X \times
SU(2)_{\rm el} \times U(1)_Y \end{split}
\eeq
where
$SU(3)_X$ is a global $SU(3)$ which will be broken by (elementary)
Yukawa couplings. The $SU(2)_{mag}\times SU(2)_{el}$ will eventually
be broken to the diagonal subgroup which will be identified with the SM
$SU(2)_L$. The embedding is chosen
so that the dual quarks contain the left-handed third generation quark
doublet, two Higgses $H_{u,d}$, and two bifundamentals ${\cal H, \bar{H}}$ that will be responsible for breaking the $SU(2)_{mag}\times SU(2)_{el}$ to the diagonal and generating the partially composite $W$ and $Z$. Fields are embedded into the dual quarks as
\beq
\begin{split} q&= Q_{3} , {\cal H},  H_d \\
\bar q&= X ,{\cal \bar{H}} ,  H_u \end{split}
\eeq
From the $q$, $\bar q$ charge assignments it follows that the meson $M$
contains the right-handed $t$, the singlets $S$ and $P$, two additional Higgses $\Phi_{u,d}$ transforming under the elementary $SU(2)_{el}$, a second right handed up-type quark $U$  and some exotics $X,V,E,R,G$:
\beq
M=\left(\begin{array}{ccc}  V & U & \bar{t} \\ E & G+P & \phi_u \\ R &
\phi_d & S \end{array} \right)
\eeq
where the quantum numbers under $SU(3)_c\times SU(2)_{el}$ for the meson fields are as follows: $V$ represents three
$(\bar{3},1)$'s, $U$ is a $(\bar{3},2)$, $E$ represents three $(1,2)$'s,
$G$ is a $(1,3)$, $\phi_d$ and $\phi_u$ are $(1,2)$'s, $P$ and $S$ are
singlets, and $R$ represents three singlets.  The
hypercharge assignments for the electric quarks, the dual quarks, and the mesons are then
\beq
\begin{array}{c}
\begin{array}{c|c|c|c|c|c|c}
& {\mathcal Q}_1 & {\mathcal Q}_2 &
{\mathcal Q}_3 & {\mathcal Q}_4& {\mathcal Q}_5 & {\mathcal Q}_6  \\
\hline Y\vphantom{Y^{Y^{Y^Y}}} \vphantom{Y_{Y_{Y_Y}}}& \frac{1}{6} & \frac{1}{6} &\frac{1}{6}& 0 &  0 & -\frac{1}{2}
\end{array}~~~,
\\ \\
\begin{array}{c|c|c|c|c|c|c|c|c|c|c|c|c|c}
\vphantom{Y^{Y^{Y^Y}}} & Q_3 & {\cal H,\bar{H}} &
H_u & H_d& X & V & U & \bar{t} & E & \phi_u & R & \phi_d & G,P,S \\
\hline Y\vphantom{Y^{Y^{Y^Y}}} & \frac{1}{6} & 0 & \frac{1}{2} &
-\frac{1}{2} &  -\frac{1}{6} & 0  & -\frac{1}{6} & -\frac{2}{3} &
\frac{1}{6} & -\frac{1}{2} & \frac{2}{3} & \frac{1}{2} & 0
\end{array}~~~.
\end{array}
\eeq
With these quantum numbers the most general  gauge invariant renormalizable electric superpotential is given by
\beq
W_{tree}= \mu_{\cal{F}} ({\cal Q}_4
\bar{{\cal Q}}_4+ {\cal Q}_5 \bar{{\cal Q}}_5) +\mu_f  {\cal Q}_6
\bar{{\cal Q}}_6 \label{Wtree}
\eeq
These will get mapped into tadpoles for the singlets $P$ and $S$ on the magnetic side. The $P$ tadpole will be responsible for the breaking of the $SU(2)_{mag}\times SU(2)_{el}$ to the diagonal, while the $S$ tadpole will be responsible for electroweak symmetry breaking.

The cancellation of SM gauge anomalies requires the presence of some  spectator fields in the electric theory that only have SM gauge couplings. A simple choice for this anomaly cancelation is  to include elementary fields that are conjugate to the representations of  composite mesons $V$, $U$, $R$, $\phi_{u,d}$, $G$.  Trilinear superpotential terms between these spectators and electric quarks will map to mass terms in the dual description, and the extra degrees of freedom will  decouple, while the fields $E,X$ will pair together to obtain a mass from the VEV of the bifundamental $\cal H$. The remaining standard model fields (first two generation quarks, right handed bottom and all leptons) are assumed to be elementary fields transforming under $SU(3)_c\times SU(2)_{el}\times U(1)_Y$. This charge assignment will be automatically anomaly free, and is capable of producing the usual flavor structure and CKM mixing matrix.


 The relevant part of the superpotential (\ref{dualsuper}) together with
the singlet tadpoles from (\ref{Wtree}) can then be written as
\beq
W\supset y P
({\cal H\bar{H}}-{\cal F}^2)  + y S (H_u H_d -f^2) + y Q_{3} H_u \bar{t}
 +  y H_u  {\cal H} \phi_u +y H_d  \bar{\cal H} \phi_d ~.
\label{dynamicalsuperpot}
\eeq
The first term is responsible
for the breaking of $SU(2)_{el}\times SU(2)_{mag}$ to the diagonal
group, the second term will trigger electroweak symmetry breaking,
the third will give rise to the $t$ Yukawa coupling and the last two
terms give rise to a mixing of the Higgs with a heavy Higgs
$\phi_{u,d}$. At this point the low-energy effective theory below the
scale ${\cal F}$ (and assuming that ${\cal F} \gg f$) is that of the
NMSSM with a composite Higgs, $Q_3$ and $t$. As explained above the rest of the SM
particles are assumed to be elementary, that is made of fields that do
not transform under the strongly coupled $SU(4)$. They simply carry the
usual SM quantum numbers under $SU(2)_{el}\times SU(3)_c\times
U(1)_Y$.

At high energies there are three sets of Higgses: the composite $H_{u,d}$ from the dual quarks transforming under the composite $SU(2)_{mag}$, the composite $\phi_{u,d}$ from the mesons transforming under the elementary $SU(2)_{el}$, and a set of elementary Higgses $\phi_{u,d}'$ transforming under the elementary $SU(2)_{el}$. These latter fields need to be present to remove $\phi_{u,d}$ from the spectrum via  a trilinear superpotential term, which after duality maps into a mass term. The elementary Higgses $\phi_{u,d}'$ also have ordinary Yukawa couplings with the light elementary SM matter fields in addition to their mass with $\phi_{u,d}$, After integrating out $\phi_{u,d}, \phi_{u,d}'$ effective Yukawa couplings between the remaining light composite Higgses $H_{u,d}$ and the light SM fermions are generated.  For more details see~\cite{MCSSM}. The resulting theory of the Higgses  in the low energy potential has the necessary Yukawa couplings and as we will now see it also has a viable and interesting potential.

\section{Electroweak Symmetry Breaking, Soft Breaking Patterns and Mass Spectrum}
\label{sec:parameters}
\setcounter{equation}{0}
\setcounter{footnote}{0}

The Higgs potential relevant for electroweak symmetry breaking (assuming
${\cal F}\gg f$) is  (including soft breaking terms)
\beq
&& V= y^2|H_u H_d  -f^2|^2   +y^2|S|^2 (|H_u|^2
+|H_d|^2) +m_S^2 |S|^2 + m_{H_u}^2 |H_u|^2 +m_{H_d}^2 |H_d|^2
\nonumber \\ &&\quad\quad+  (A S H_u H_d   + T S+ h.c.)+
\frac{g^2+g'^2}{8} (|H_u|^2 -|H_d|^2 )^2 \label{Higgspot}
\eeq
where $m_{S,H_u,H_d}^2$, $A$ and $T$ are soft supersymmetry breaking
parameters, and the last term is the usual MSSM $D$-term.  This is quite
different from the usual MSSM potential, and the traditional source of
fine tuning related to the need of large ${\tilde t}$ loop corrections for the
quartic are not produced.  While the matter content of the Higgs sector is that of an NMSSM,
the actual potential is quite different from what is traditionally used in a $Z_3$ symmetric NMSSM. Electroweak symmetry is broken in the supersymmetric limit, and a Higgs mass much bigger than in the MSSM is ensured since the quartic does not come from $D$-terms and thus the Higgs mass is not related to the $Z$-mass. Such
Higgs sectors are natural in the context of composite ``fat Higgs"-like
models \cite{fatHiggs1,fatHiggs2}: the NMSSM singlet $S$ is simply one
of the composite meson components. The NMSSM-like superpotential
given in Eq. (\ref{dynamicalsuperpot})  is the one that appears
most naturally in Seiberg duals. The electroweak symmetry breaking scale is determined by the magnitude of the $S$-tadpole $f$, which means that electroweak symmetry breaking
 in general is not dependent or related to supersymmetry breaking, but that $f$ has to be of the order of the Higgs VEV $v$. For a completetly natural model, one would hope for a deeper relation between $f$ and $v$.   This is similar to the usual $\mu$-problem of the MSSM (without a corresponding $B\mu$ problem). The traditional way of solving this would be to assume that the electric theory has a global Peccei-Quinn-type symmetry that forbids the mass term for the electric quarks that  eventually turn into the composite $S$, and that this PQ symmetry is only broken in the supersymmetry breaking sector. Coupling the electric quarks to the supersymmetry breaking sector can then give a PQ violating superpotential term proportional to the supersymmetry breaking scale just like in the usual Giudice-Masiero mechanism. We will not try to build a complete model for the supersymmtry breaking sector in this paper.

We will use the usual parametrization of the Higgs fields:
\beq
H_u=\left( \begin{array}{c} H^+_u \\ H^0_u \end{array} \right), \quad \quad   H_d=\left( \begin{array}{c} H^0_d \\ H^-_d \end{array} \right)  \\
\langle H_u^0
\rangle = \frac{v}{\sqrt{2}} \sin \beta~,\quad \quad   \langle H_d^0 \rangle =
\frac{v}{\sqrt{2}} \cos \beta ~.
\eeq
Since the  interaction with the singlet provides a sizable
quartic, it is not important to have a large $\tan \beta$, it actually
can be close to one, or even less than one.  Minimizing the potential with respect to the scalar
$S$ we find the scalar VEV
\beq
\langle S\rangle = -\frac{\sqrt{2}
\left(\text{A} v^2 \sin \beta \cos \beta +2 T\right)}{2 M_S^2+y^2 v^2
}~, \label{SVEV}
\eeq
A combination of the other two equations  yield an expression that is analogous to the usual
fine-tuning condition for the Higgs VEV:
\beq
\frac{y^2 v^2}{2} =
\frac{2(y^2 f^2 -A S)}{\sin 2\beta} -2 y^2 S^2 -m_{H_u}^2 -m_{H_d}^2
\eeq
Thus the fine tuning can now be characterized by
\beq
\frac{y^2 v^2}{2m_{H_u}^2}
\label{eq:finetuning}
\eeq
In most supersymmetric models, the $\tilde{t}$'s  have to be sufficiently heavy  to generate a large enough  Higgs quartic (or equivalently, a large enough physical Higgs mass). On the other hand, heavy $\tilde{t}$'s also give a large contribution to $m_{H_u}^2$ leading to large tuning. In our models, one has a large tree-level quartic from compositeness, and the ${\tilde t}$'s are light, thus (\ref{eq:finetuning}) can be of ${\cal O}$(1) with composite ${\tilde t}$ masses  in the 200-500 GeV range.  Even so, since the gluino is elementary and thus in the few TeV range  the the two-loop corrections to the Higgs mass via gluino-${\tilde t}$ loops can potentially be too large.
The leading 2-loop correction to $m_{H_u}^2$ due to the gluino loop is
\beq
\Delta m_{H_u}^2 \sim -\frac{2 y_t^2 \alpha_s^2}{\pi^3}
|m_{\tilde{g}}|^2 \log^2 \left( \frac{\Lambda}{{\rm TeV}} \right)
\eeq
Note
that due to compositeness, the cutoff scale of the logarithm is small
here. Even for low $\tan \beta$,  one gets only about ten percent
tuning for a gluino as heavy as 3 TeV.

We conclude that in principle, a gluino heavier than those that are usually considered natural would be allowed.
However, a heavy gluino mass would also contribute to the ${\tilde t}$ masses, and in our models we assume light top squark masses. The leading log correction to the ${\tilde t}$ mass parameters is of the order
\begin{equation}
\Delta m_{\tilde{t}} \sim \frac{32}{3} \frac{\alpha_s}{4\pi} |M_3|^2 \log \left(  \frac{\Lambda}{\rm TeV} \right)
\end{equation}
Even with this additional consideration on naturalness, since the logarithm is quite small (corresponding to the running between the duality scale and the TeV scale, $\log \frac{\Lambda}{\rm TeV} \sim 2$), one can naturally maintain a  hierarchy between the gluino and the ${\tilde t}$ mass. However this hierarchy cannot be very large if we want to keep the top squark light.
A gluino of about 1.5 TeV would be natural with a 400 GeV ${\tilde t}$ without much tuning. If one were to allow ten percent tuning the gluino mass could be raised to about 3 TeV. We will however not do that, and restrict the gluino mass to be below 1.5 TeV in order to protect the squark mass hierarchies obtained from the strong dynamics. Note, that the experimental lower bound on the gluino is around 700 GeV even if it only decays via third generation squarks~\cite{Matt}.

 We now discuss the pattern of soft breaking terms and the magnitudes
of the relevant parameters of the model. While we do not
fully specify the mechanism of supersymmetry breaking mediation to the
elementary (``electric") fields here, we will usually assume some
form of low-scale mediation mechanism, in order to have the gravitino be
the LSP. The prime example of such models is gauge mediation. However, even if we assume gauge mediation applies,
this is a non-standard application, since we are
eventually ending up with the NMSSM. Naively one would think that gauge
mediation can not be applied to an NMSSM-type theory, since the singlet
will not obtain SUSY breaking terms. However, in this case gauge
mediation is assumed to happen above the compositeness (``duality")
scale. Since the singlet is a composite (it is a component of the meson) a soft breaking term (suppressed as with all composites) will be induced for it. The mass for the fermionic partner of the singlet (the singlino) is model dependent. There can be a singlino mass from non-renormalizable terms for the elementary fields $(\bar{\cal Q}_6 {\cal Q}_6)^2/\Lambda_{UV}$ giving a singlino mass of order $m_{S_f}\sim \Lambda^2/\Lambda_{UV}$. There will also be a singlino mass generated by the strong dynamics of order $\frac{f^4}{\Lambda^4} m_{el}$ which is typically quite small. We will not be making a definite assumption on the size of the singlino mass, but explore spectra both with small and sizeable values for it.

Note that the usual  $B \mu$ problem  is simply not  present, since the potential contains only trilinear and tadpole terms, both of which
are induced as described in Sec.~\ref{sec:soft}.  While the $\mu$-problem is solved as usual in NMSSM-type models, an issue similar to the $\mu$-problem is why the parameter $f$ is close to the electroweak scale, which as we discussed before is likely to be addressed with a more complete model of SUSY breaking.

 The message from the
general discussion of Section \ref{sec:soft} is that soft breaking terms for the composites
are suppressed compared to those of the elementary
fields, while the scalar tadpole $T$ is unsuppressed. We  choose parameters
consistent with the hierarchies explained in the previous explained in the previous section of order
\beq
&& m_{el} \sim M_3 \sim
{\rm few}\cdot {\rm TeV} \nonumber \\ && \Lambda \sim 5-10\  {\rm TeV}
\nonumber \\ &&m_{comp}\sim \frac{m_{el}^2}{\Lambda} \sim M_1 \sim M_2 \sim A
\sim {\rm few}\cdot 100 \ {\rm GeV} \nonumber \\ && f\sim 100\  {\rm
GeV} \nonumber \\  && T\sim f^2 m_{el} \sim {\rm few} \cdot
10^7 \ {\rm GeV}^3 \nonumber \\ && {\cal F} \sim {\rm few} \cdot\  {\rm
TeV} \nonumber \\ && \mu_{\rm eff} = y \langle S \rangle \sim A \\
&& \tan \beta \sim {\cal O}(1)
\eeq

Here $m_{el}$ includes
the soft breaking scalar masses of the first two generation squarks, the
right handed sbottom, ${\tilde b}$ and all sleptons, while $m_{comp}$ includes
$m_{Q_{33}}$ and $m_{U_{33}}$. The soft terms include the dynamical non-calculable contributions of ${\cal O}(m_{el}^2/\Lambda)$ and the additional radiative corrections $\propto \log \frac{\Lambda}{\rm TeV}$. The latter can be comparable to the dynamical terms as we discussed for the gluino loops. The effective $B\mu$ term is
$A\langle S\rangle \sim \mu_{\rm eff}^2$. However, as stated previously,  in this model
electroweak symmetry is broken in the supersymmetric limit, so the
magnitude of $B\mu$ is not very crucial. Note, that flavor constraints for such models with heavy first and second generation squarks and sleptons are largely satisfied if the scale of the heavy squark masses is around 5 TeV~\cite{flavorconstraints}, and if the heavy squarks are close to degenerate, which would be the case if they get their masses from gauge mediation.

With this choice of parameters we can then go ahead and evaluate the
full sparticle spectrum.  We present the relevant expressions for the masses
below, while in the next section we focus on four benchmark
spectra.

Given all the soft SUSY breaking terms the spectrum calculation proceeds
in a similar fashion to the MSSM and NMSSM. The ${\tilde t}$ mass matrix is
\beq
m_{\tilde t}^2 = \left(\begin{array}{cc} m^2_{Q33} + m_t^2
+\delta_{u} & v(A \, s_\beta -\mu_{\rm eff}\, y_t\,c_\beta )/\sqrt{2}\\
v (A \, s_\beta - \mu_{\rm eff}\, y_t\,c_\beta )/\sqrt{2} &
m^2_{\overline u 33} + m_t^2 + \delta_{\overline u} \end{array}\right) ,
\eeq
where the D-term contribution is as usual
\beq
\delta_f=-g T^3_f
\langle D^3\rangle - g^{\prime}Y_f \langle D^\prime\rangle =(T^3_f - Q_f
s^2_W) \cos 2\beta\, M_Z^2~.
\eeq
Since the ${\tilde b}$ mass is constrained by the LHC to be above $\sim 250 -280$ GeV~\cite{lightstopAndi},
$m_{Q33}$ should not be too small, since this sets the mass of the lighter ${\tilde b}$. The right-handed ${\tilde t}$ mass, $m_{\bar{u}33}$, can be somewhat smaller than $m_{Q33}$, and with $A$ not too large one gets a spectrum with the right handed ${\tilde t}$ as the lightest sfermion, a somewhat heavier left handed ${\tilde t}$ and left handed ${\tilde b}$, while the elementary fields are quite a bit heavier.

The explicit form of the ${\tilde b}$ mass matrix is
\beq
{ m_{\tilde b}^2} = \left( \begin{array}{cc} m^2_{Q33} +m_b^2 +
\delta_{d} & v (A_{d33}\, c_\beta -\mu_{\rm eff}\, y_b\,s_\beta
)/\sqrt{2}\\ v (A_{d33}\, c_\beta - \mu_{\rm eff}\, y_b\,s_\beta
)/\sqrt{2} & m^2_{\overline d33} +m_b^2 + \delta_{\overline d}
\end{array}\right)~,
\label{msbottommatrix}
\eeq
where the right handed ${\tilde b}$ is elementary, so its soft breaking mass is
expected to be large $m^2_{\overline d33}\sim m_{el}$,  while
$m^2_{Q33}\sim m_{comp}$ is suppressed.

Due to the extra $SU(2)$ group we have an additional set of charginos
and neutralinos, and the singlet $S$ also contributes to the neutralino
mass matrix. The chargino mass matrix  is
\beq
\left(
\begin{array}{cccc} \widetilde{W}_{2,el}^{-} &  \widetilde{\phi}_d^- & \widetilde{H}_d^- &
\widetilde{W}_{2,mag}^{-} \end{array} \right) \left(\begin{array}{cccc} M_2
& \frac{g_{el}}{\sqrt{2}}{\mathcal F}  & 0 & 0 \\ \frac{g_{el}}{\sqrt{2}}
{\mathcal F}   & y\langle P\rangle &  0 & \frac{g_{mag}}{\sqrt{2}}{\mathcal
F} \\ 0&   0 &\mu_{\rm eff}\, & \frac{g_{mag}}{\sqrt{2}} s_\beta \,v  \\ 0
& \frac{g_{mag}}{\sqrt{2}}{\mathcal F}   & \frac{g_{mag}}{\sqrt{2}} c_\beta \,v
  &M_{2,mag} \end{array} \right)~\left( \begin{array}{c} \widetilde{W}_{2,el}^{+} \\  \widetilde{\phi}_u^+ \\ \widetilde{H}_u^+ \\ \widetilde{W}_{2,mag}^{+}
\end{array}
\right)
\label{chargino}
\eeq
where we have also added the elementary winos for the $SU(2)_{el}$ group and the higgsinos from $\phi_{u,d}$, and also ${\mathcal F}=\langle {\mathcal
H}\rangle$ is the bifundamental VEV that breaks the composite
$SU(2)_{\rm mag}$ and the elementary $SU(2)_{\rm el}$ down to the
diagonal $SU(2)_L$ subgroup, while  $g_c$ and $g_e$ represent the two
gauge couplings giving rise to the SM couplings via the mixing
\beq
s_\theta = \frac{g_{el}}{\sqrt{g_{el}^2+g_{mag}^2}}~, \quad\quad  c_\theta =
\frac{g_{mag}}{\sqrt{g_{el}^2+g_{mag}^2}}~,\quad\quad g_2= g_{mag} s_\theta = g_{el}
c_\theta~.
\eeq
and $\mu_{eff}=y \langle S\rangle$.
In the limit $g_{mag} \gg g_{el}, {\mathcal F}\gg
M_2, M_{2,mag}$ this can be
approximately diagonalized,
and the heavy
combination of gauginos (corresponding mostly to the composite charginos
and the ${\cal H}$'s)  can be integrated out with only small corrections
to the mass spectrum that results from the ordinary MSSM mass matrix of
the form
\beq
\left( \begin{array}{cc} \widetilde{H}_d^- & \widetilde{W}_{2,L}^{-} \end{array} \right) \left(
\begin{array}{cc} \mu_{\rm eff}\, &
\frac{g_2}{\sqrt{2}} s_\beta \,v  \\ \frac{g_2}{\sqrt{2}} c_\beta \,v
& M_2 \end{array} \right)~\left( \begin{array}{c}   \widetilde{H}_u^+ \\
\widetilde{W}_{2,L}^{+} \end{array} \right)
\label{mssmchargino}
\eeq
with the
elementary gaugino mass playing approximately  the role of the
MSSM $M_2$ parameter. In some regions of parameter space the extra
mixing can change the chargino spectrum but we will not consider that
case here. Thus to leading order it is the elementary gaugino that will be lighter, due to the large coupling of the composite gaugino. When gauginos are light,  it is as usual only because of the suppression by the small SM gauge couplings.

Similarly the neutralino mass matrix reduces to the NMSSM
form after integrating out the heavy neutral fermions corresponding to
the composite neutral gauginos:
\beq
\left( \begin{array}{ccccc} M_1 & 0
& -M_Z c_\beta s_W & M_Z s_\beta s_W & 0 \\ 0 & M_2 & M_Z c_\beta c_W &
-M_Z c_W s_\beta & 0 \\ -M_Z c_\beta s_W & M_Z s_\beta s_W & 0 &
-\mu_{\rm eff}  & -y v s_\beta \\ M_Z c_\beta c_W & -M_Z c_W s_\beta &
-\mu_{\rm eff} & 0 & -y v c_\beta \\ 0 & 0 & -y v s_\beta & -y v c_\beta
& M_{Sf} \end{array} \right)~.
\eeq
where $s_W=\sin \theta_W$, $c_W=\cos
\theta_W$, where again $M_{1,2}$ are approximately given by the
elementary gaugino masses, and we have also included a soft breaking
Majorana mass for the singlino.  All other fields either correspond to
elementary fields with large SUSY breaking terms, or are vector-like and
also assumed to have large masses. This way we obtain the particle
spectrum we will be investigating in the next section: the lightest ${\tilde t}$
within a few hundred GeV of the top mass, heavier ${\tilde t}$ and lighter
${\tilde b}$ below 500 GeV, neutralinos and charginos and the full scalar
Higgs sector below a TeV, while all other particles are above one TeV.

\section{Phenomenology of a Light Composite Stop}
\label{sec:stealth}
\setcounter{equation}{0}
\setcounter{footnote}{0}

Finally we discuss the phenomenology of composite supersymmetric models
with light ${\tilde t}$'s. We restrict out analysis to regions of parameter space
for which  the lighter ${\tilde t}$ (which
is mostly the right handed ${\tilde t}$) is within a few hundred GeV of  the
top.  We  examine four different spectra in order to display a variety of phenomenological possibilities.

The NLSP will be either the ${\tilde t}$ or the lightest neutralino, $N_1$. The first two spectra have ${\tilde t}$ (N)LSP. The two spectra are distinguished by the degree of degeneracy of the $t$ and right handed ${\tilde t}$. In  the first, the ${\tilde t}$  NLSP is nearly degenerate with the $t$, generating a stealth stop spectrum, while  the ${\tilde t}$ is a bit heavier for the second parameter set. The third spectrum has a neutralino NSLP and supersymmetry breaking arises form standard gauge mediation.   The last spectrum has a neutralino (N)LSP, but the dominant contribution to the soft mass parameters is assumed to be the radiative contributions, and not the power suppressed corrections. In other words, this model assumes a relatively high compositeness scale.

 When the NSLP is the ${\tilde t}$, it will decay to $t$ plus gravitino. If it is the $N_1$, then  the ${\tilde t}$ will decay (depending on kinematics) either to $t +N_1$ or bottom plus chargino ($b+C^-$), while the $N_1$ will decay to photon plus gravitino or Higgs/$Z$+gravitino.   Alternatively the $N_1$ may be the LSP itself, with higher scale SUSY breaking and heavier gravitino.  In either case there will be missing energy signals from neutralino production.

For the spectra where the ${\tilde t}$ is lighter than the $N_1$, we  assume a low-scale for supersymmetry breaking
$\sqrt{F} \leq 10^{10}$ GeV, implying an LSP gravitino mass  of a
few GeV or less. As long as the mediation scale $M_{SUSY}$ is well above
the duality scale $\Lambda \sim 5-10 $ TeV the assumption that
supersymmetry breaking must be fed through the duality applies.

The viability of a $t$-${\tilde t}$ sector with a ${\tilde t}$ NLSP decaying via the gravitino has recently been investigated
in detail by Kats and Shih in~\cite{Kats} using Tevatron and first year
LHC data (35 pb$^{-1}$). They found that using searches based on these
data sets  that the  data on the lightest ${\tilde t}$ mass decaying to $t$
plus gravitino sets a bound of about ${\tilde t}$ mass of  about 150  GeV, and that bounds of about 180
GeV are expected using 3 fb$^{-1}$ data.  If the lightest ${\tilde t}$ mass is
almost degenerate with the top, then there will not be much missing
energy in the decays leading to the stealth supersymmetry scenario
mentioned in~\cite{Fan}. The most recent papers~\cite{lightstopGMSB,lightstopRaman,lightstopAndi,lightstopSLAC} on
light third generation bounds from 1 fb$^{-1}$ of LHC data have also considered the possibility of the lightest $\tilde{t}$ decaying to top plus gravitino. They have found (in agreement with~\cite{Kats}) that currently there is no bound~\cite{lightstopAndi} over 200 GeV for such a $\tilde{t}$.

These most recent analyses~\cite{lightstopGMSB,lightstopRaman,lightstopAndi,lightstopSLAC} have also examined bounds on the heavier $\tilde{t}$ and the left handed $\tilde{b}$.  These are assumed to decay to neutralinos/charginos, and for decays of this type the currents bounds are found to be around 270 GeV.  We
take it as an indication that left handed ${\tilde t}$'s and ${\tilde b}$'s of order
300 GeV are experimentally viable, even though in some of the spectra presented here the leading decays of the heavier $\tilde{t},\tilde{b}$ will actually involve the lighter $\tilde{t}$.

We now discuss our choice of input parameters that correspond to these spectra. When minimizing the
Higgs potential (\ref{Higgspot}), we impose the EWSB vacuum with the
correct value of $v$ and a fixed $\tan\beta$, with an appropriate choice of the scalar
tadpole $f$. This fixes the values of the Higgs soft breaking
terms $m_{H_{u,d}}^2$, which will not be treated as inputs. We do
however check that these  terms have the correct magnitudes presented in the previous section.
The other relevant input parameters  to fix are the composite
soft breaking masses $m_{Q_{33}}, m_{u_{33}}, M_{1,2}$ and $M_S$.
As discussed before, a  Majorana mass for the singlet fermion
$M_{S_f}$ may also be present, and in the second spectrum we add a term that raises the neutralino mass. In all other spectra this term is set to zero.  The $A$-terms for the $SH_uH_d$ and the $t$ Yukawa interaction
originate from the same dynamical term and are thus assumed to be equal. Finally we need to asign the soft breaking scalar tadpole $T$.
All other
soft breaking masses are assumed to be above a TeV, ensuring
that the rest of the sparticle spectrum is essentially decoupled due to
them being elementary degrees of freedom. Elementary Higgses responsible for generating the Yukawa couplings for the elementary fields are assumed to be heavy and integrated out for the purposes of this paper, but it could be interesting to investigate a theory with the elementary Higgses included as light fields as well.

The input parameters for the four benchmark spectra are given in Table \ref{input}. Minimizing (\ref{Higgspot}) and imposing the correct electroweak symmetry breaking VEV's fixes $\mu_{\rm eff}$, $m_{H_u}^2$,  $m_{H_d}^2$; the corresponding values  are given in Table \ref{output}. The first two spectra we  examine have ${\tilde t}$ NLSP's, while the second two have neutralino NLSP/LSP's. The singlino mass is set to zero in all but the second spectrum, where it is used to raise the lightest neutralino mass above the ${\tilde t}$ mass. The first spectrum has the lightest ${\tilde t}$ almost degenerate with the $t$, and is thus more ``stealthy", while the second one has heavier ${\tilde t}$'s with it still being the NLSP.  The third spectrum implements minimal gauge mediation to the electric degrees of freedom: the ratio of gaugino masses here is given by the coupling constant squares (with the gluino at 1 TeV), and the other soft breaking masses for the composites taken equal. The fourth spectrum was chosen such that the soft-breaking Higgs masses are rather small so this scenario could correspond to a high duality scale with radiatively generated $\tilde t$ and $\tilde b$ masses. While we are assuming some form of low-scale supersymmetry breaking in all but one  of the spectra, only the third one corresponds to minimal gauge mediation. In the minimal case the gaugino mass ratios are determined  by the SM gauge couplings, and the upper bound on the gluino mass implies a fairly light bino below 100 GeV and thus a neutralino LSP (unless the a large contribution to the singlino mass is present). The cases with heavier gaugino masses (and ${\tilde t}$ NLSP's) can be thought of as cases corresponding to a general gauge mediated spectrum~\cite{GGM} to the electric degrees of freedom.

\begin{table}
\begin{center}
$
\begin{array}{c|c|c|c|c}
{\rm
parameter}&	 {\rm spectrum}\,\,\, 1		& {\rm spectrum}\,\,\, 2 		& {\rm spectrum}\,\,\, 3		& {\rm spectrum}\,\,\, 4	\\ \hline
 \tan \beta	&    		0.85 					&	 1.3 					& 1.0					& 0.97 					 \\
 A		& 300 \,\,\,{\rm GeV}				& 540 \,\,\,{\rm GeV}			&	350 \,\,\,{\rm GeV}		& 400 \,\,\,{\rm GeV}	\\
 T		& 4 \times 10^7 \,\,\,{\rm GeV}^3	& 1.4 \times 10^7 \,\,\,{\rm GeV}^3 & 3.35 \times 10^7 \,\,\,{\rm GeV}^3 & 6 \times 10^6 \,\,\,{\rm GeV}^3 	\\
m_{Q_{33}}	&  500 \,\,\,{\rm GeV}		& 500 \,\,\,{\rm GeV}               & 350 \,\,\,{\rm GeV}                          & 400 \,\,\,{\rm GeV} \\
m_{U_{33}}	&  250 \,\,\,{\rm GeV}	 	& 350 \,\,\,{\rm GeV}               & 350 \,\,\,{\rm GeV}                & 400 \,\,\,{\rm GeV}	\\
M_1		& 600 \,\,\,{\rm GeV}				& 700 \,\,\,{\rm GeV}			& 85 \,\,\,{\rm GeV}			& 600 \,\,\,{\rm GeV}	 \\
M_2		& 800 \,\,\,{\rm GeV}				& 800 \,\,\,{\rm GeV}		 	& 282 \,\,\,{\rm GeV} 		& 1200 \,\,\,{\rm GeV} \\
m_S	  & 400 \,\,\,{\rm GeV}					& 350\,\,\,{\rm GeV} & 350 \,\,\,{\rm GeV}                             & 100 \,\,\,{\rm GeV} \\
M_{Sf}	& 0 \,\,\,{\rm GeV}				& -350\,\,\,{\rm GeV}                   & 0 \,\,\,{\rm GeV}                    & 0 \,\,\,{\rm GeV}\\
f 		& 100 \,\,\,{\rm GeV} 			&  100 \,\,\,{\rm GeV} 		&  293 \,\,\,{\rm GeV}		&  100 \,\,\,{\rm GeV}
\end{array}
$
\caption{Input parameters for the four sample spectra. In spectrum 1, the ${\tilde t}$ is the NLSP and very degenerate with the top, generating a stealth stop spectrum. In spectrum 2, the ${\tilde t}$ is the NLSP but is a bit heavier. Spectrum 3 has a neutralino NLSP and is generated through a gauge mediated spectrum. Spectrum 4 has a neutralino (N)LSP, and the compositeness scale is assumed high enough that radiative corrections to soft composite superpartners dominate.\label{input}}
\end{center}
\end{table}

We have chosen the parameters of all four spectra such that the lightest Higgs mass is around 125 GeV. This is not a necessity dictated by the model, and one can easily obtain spectra with heavier Higgses. We also made sure that for these points we are sufficiently close to the decoupling limit, such that Higgs production and decay rates are not too far from the corresponding SM values.
Note that choosing the input parameters given above does not involve any extensive
tuning: no automated scans had to be performed for finding these points.

\begin{table}
\begin{center}
$
\begin{array}{c|c|c|c|c}
{\rm
parameter}&	 {\rm spectrum}\,\,\, 1		& {\rm spectrum}\,\,\, 2 			& {\rm spectrum}\,\,\, 3		& {\rm spectrum}\,\,\, 4	\\ \hline
\mu_{\rm eff}& -416  \,\,\,{\rm GeV}			& -639\,\,\,{\rm GeV}				& -422 \,\,\,{\rm GeV}		& -342 \,\,\,{\rm GeV}\\
m_{H_u}^2& -(176 \,\,\,{\rm GeV})^2		& -(244 \,\,\,{\rm GeV})^2			& (350 \,\,\,{\rm GeV})^2		& (40.3 \,\,\,{\rm GeV})^2\\
m_{H_d}^2& -(218 \,\,\,{\rm GeV})^2		& (207 \,\,\,{\rm GeV})^2			& (350 \,\,\,{\rm GeV})^2		& -(46.6 \,\,\,{\rm GeV})^2
\end{array}
$
\caption{Output parameters for the four benchmark spectra.\label{output}}
\end{center}
\end{table}

In order to calculate  the spectrum and widths we have modified the
NMSSMTools \cite{NMToolscode,NMTools} package, which deals with the
$Z_3$ symmetric NMSSM. The modified package (MCSSMTools)
\cite{MCSSMTools}   handles the minimal composite supersymmetric
standard model considered here, where a linear superpotential term,
tadpole soft breaking term, and a singlino mass are also allowed.

\begin{figure} \begin{center} \begin{tabular}{cc}
\includegraphics[height=9.4cm]{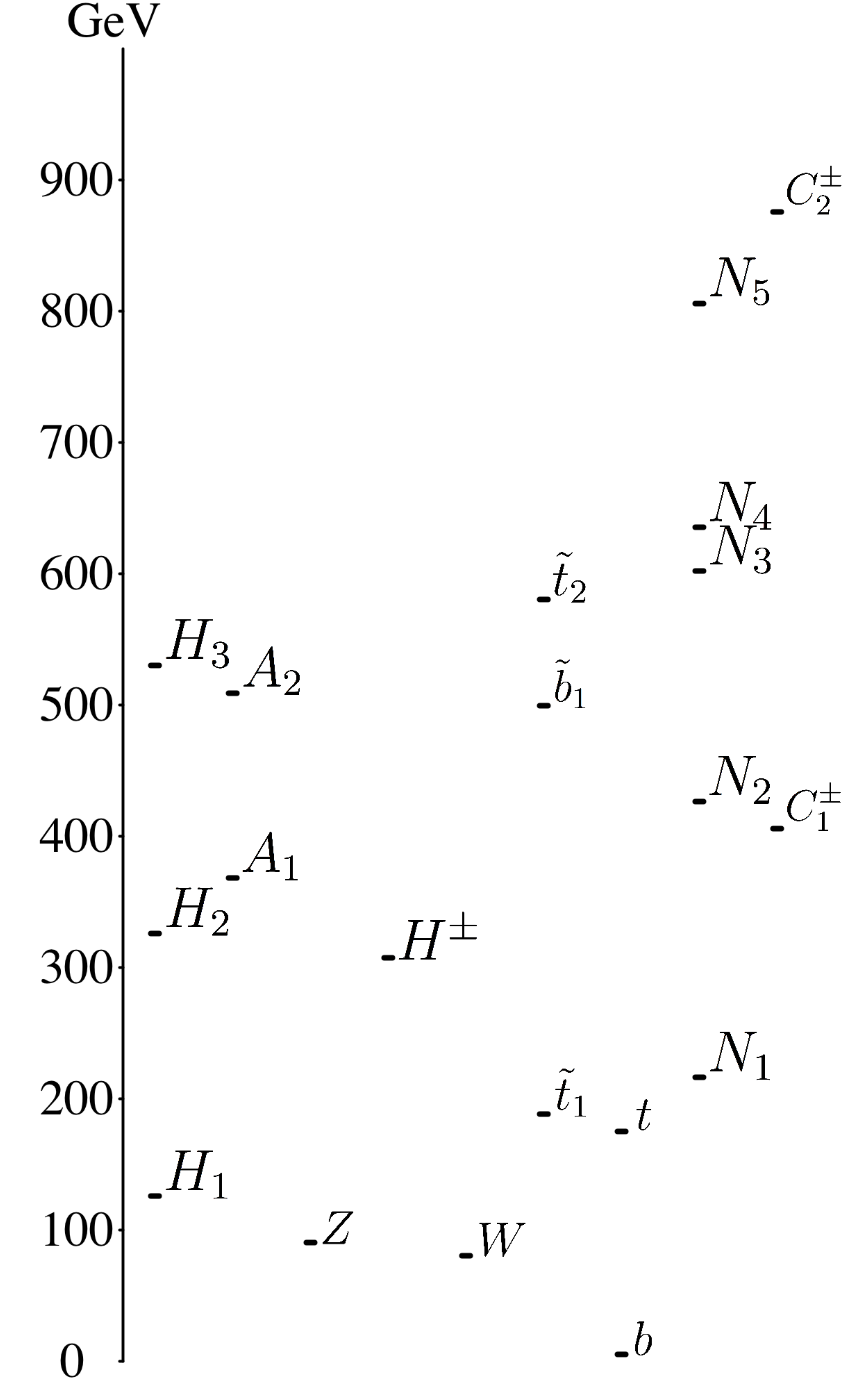}\hspace{0.9cm}\includegraphics[
height=9.4cm]{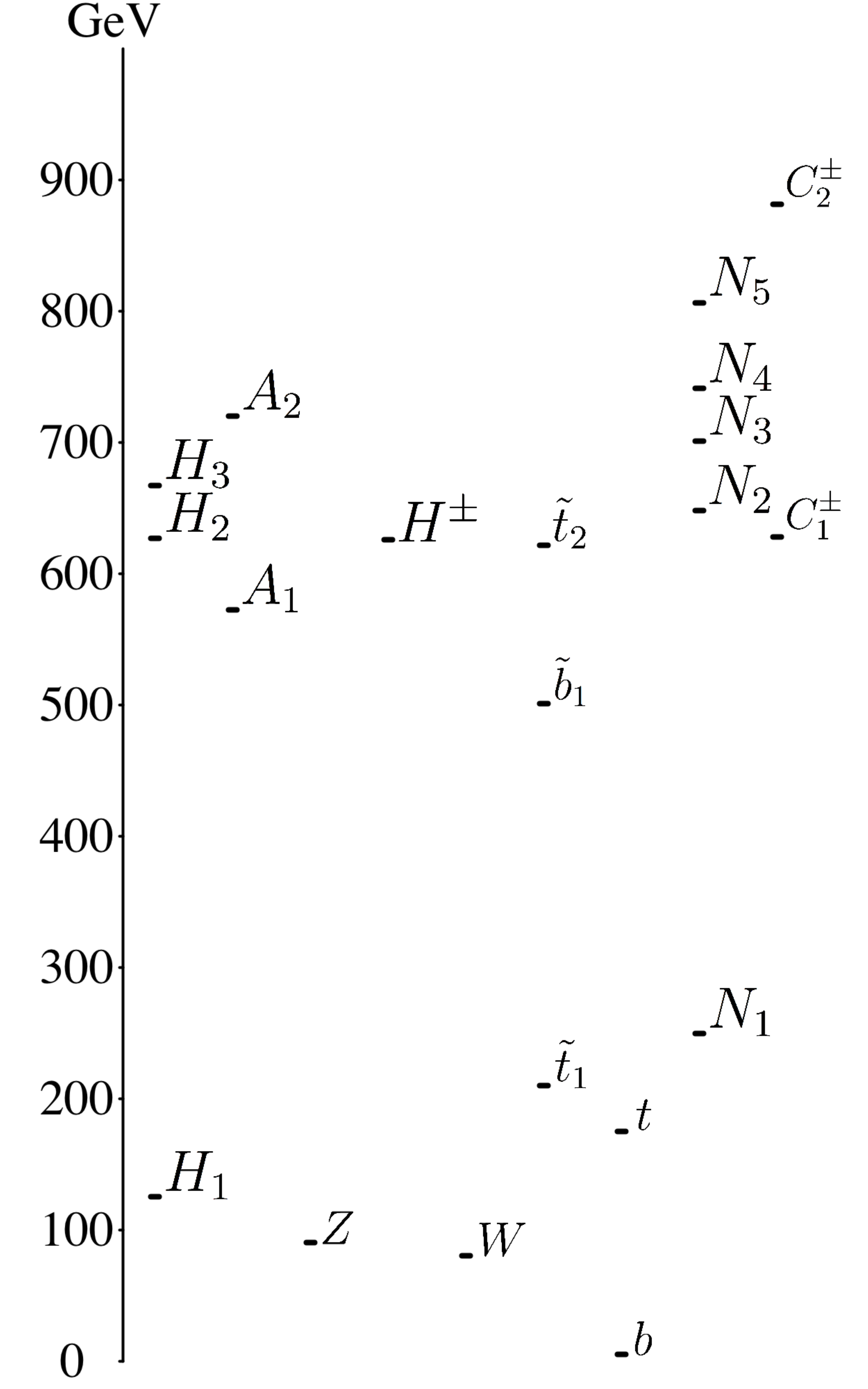}
\end{tabular} \vspace*{-0mm} \end{center}
\caption{Light superpartners and Higgs particles for benchmark  spectra 1 and 2 with a ${\tilde t}$ NLSP.}
\label{spectrum12}
\end{figure}

\begin{figure} \begin{center} \begin{tabular}{cc}
\includegraphics[height=9.4cm]{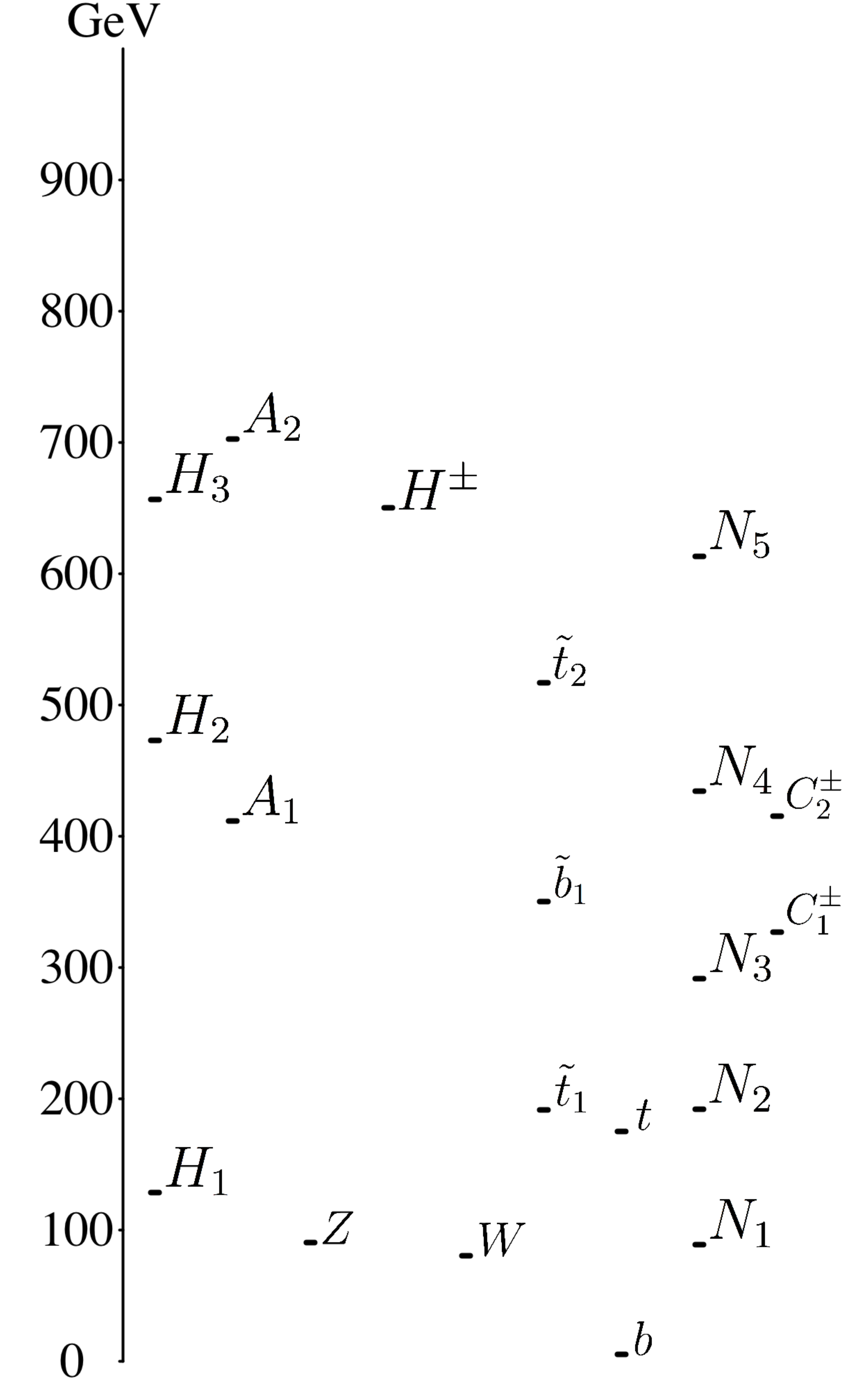}\hspace{0.9cm}\includegraphics[
height=9.4cm]{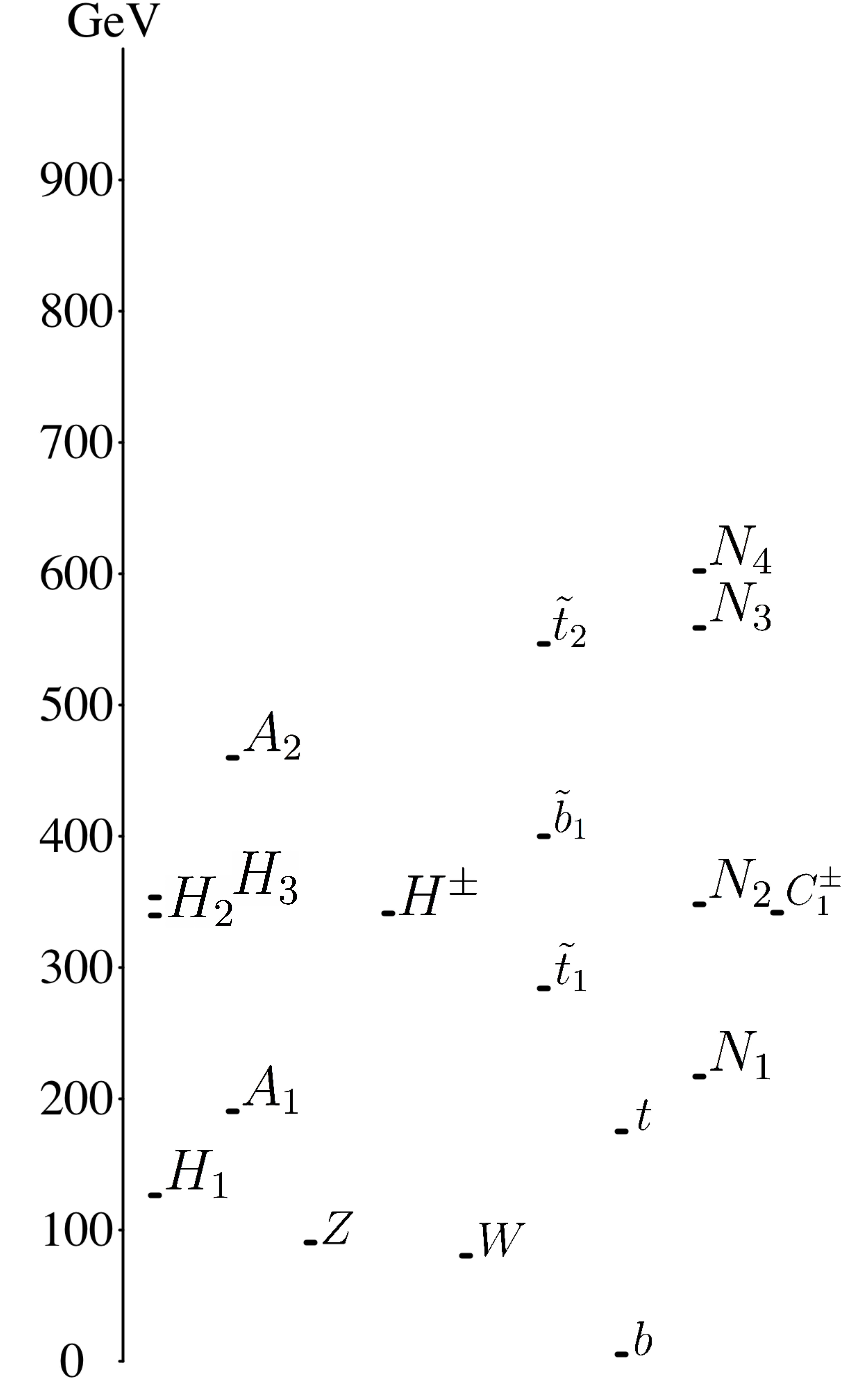} \end{tabular} \vspace*{-0mm} \end{center}
\caption{Light superpartners and Higgs particles for benchmark  spectra 3 and 4.}
\label{spectrum34}
\end{figure}

The mass spectra are presented graphically in Fig. \ref{spectrum12} (benchmark spectra 1 and 2 with ${\tilde t}$ NLSP's) and
Fig. \ref{spectrum34} (benchmark spectra 3 and 4 with neutralino NLSP/LSP's). The numerical values for the masses for spectra 1 and 2 are presented in
Table \ref{masses12}, while the leading decay modes are in  Table \ref{decay12}.
The physical masses for spectra 3 and 4 are in Table \ref{masses34}, with decay modes in Table \ref{decay34}.
The spectrum and decay chains can be interactively visualized online at
\href{http://bit.ly/mcspect}{\tt http://bit.ly/mcspect}.  Table \ref{Higgscouplings} contains the couplings of the lightest Higgs relative to their SM values. One can see that we are close to the decoupling limit in each case: gluon couplings are within 65-83\% of the SM values, while the photon coupling varies between 85-102\% of the SM size for the same Higgs mass.

\begin{table}
\begin{center}
\begin{tabular}{cc}
$
\begin{array}{|c|c|c|c|}
\hline H_1  & 125      \,\,\,{\rm GeV} &  \tilde b_1  \vphantom{\sqrt{\tilde b}}  & 499 \,\,\,{\rm GeV}\\
\tilde t_1 & 188\,\,\,{\rm GeV}&  A_2 & 509  \,\,\,{\rm GeV} \\
N_1 & 216      \,\,\,{\rm GeV}  & H_3 & 530  \,\,\,{\rm GeV}  \\
H^\pm & 307   \,\,\,{\rm GeV}&  \tilde t_2   & 580 \,\,\,{\rm GeV} \\
H_2 & 326    \,\,\,{\rm GeV}& N_3 & 602   \,\,\,{\rm GeV} \\
A_1 & 368    \,\,\,{\rm GeV}& N_4 & 635  \,\,\,{\rm GeV} \\
C_1   & 406     \,\,\,{\rm GeV}& N_5 & 805   \,\,\,{\rm GeV} \\
N_2 & 426  \,\,\,{\rm GeV}& C_2 & 876  \,\,\,{\rm GeV} \\ \hline
\end{array}
$\quad \quad\quad& \quad\quad\quad
$
\begin{array}{|c|c|c|c|}
\hline
H_1 & 125   \,\,\, {\rm GeV}   	& C_1 & 628    \,\,\, {\rm GeV}    \\
\tilde t_1 & 210  \,\,\, {\rm GeV}  & N_2 & 651  \,\,\, {\rm GeV} \\
N_1 & 429  \,\,\, {\rm GeV} &  H_3 & 667 \,\,\, {\rm GeV}  \\
\tilde b_1 & 501  \,\,\, {\rm GeV}  & N_3 & 700 \,\,\, {\rm GeV}  \\
A_1 & 572  \,\,\, {\rm GeV}  	&  A_2 & 720  \,\,\, {\rm GeV}  \\
\tilde t_2  & 621 \,\,\, {\rm GeV} & N_4 	& 724    \,\,\, {\rm GeV}  \\
H^\pm & 626 \,\,\, {\rm GeV} & N_5 & 806  \,\,\, {\rm GeV}  \\
H_2 & 627  \,\,\, {\rm GeV} &  C_2 & 881  \,\,\, {\rm GeV}   \\ \hline
\end{array}
$
\end{tabular}
\end{center}
\caption{Light superpartners and Higgs
particles for benchmark  spectra 1 and 2 with a ${\tilde t}$ NLSP. All other superpartners are above 1 TeV. \label{masses12}}
\end{table}

\begin{table}
\begin{center}
\begin{tabular}{cc}
$
\begin{array}{clc} {\tilde t}_1 & \rightarrow t +LSP& 100\% \\
C_1 & \rightarrow {\tilde t}_1 +b^\dagger & 84\% \\
C_1 & \rightarrow N_1 +W^\pm  & 16\% \\
{\tilde b}_1 & \rightarrow {\tilde t_1} +W^- & 97\% \\
{\tilde b}_1 & \rightarrow {\tilde t_1} +H^- & 3\% \\
{\tilde t}_2 & \rightarrow {\tilde t_1} +Z& 51\% \\
{\tilde t}_2 & \rightarrow t +N_1 & 27\%  \\
{\tilde t}_2 & \rightarrow b+ C_1^+ & 11\%  \\
{\tilde t}_2 & \rightarrow {\tilde t_1}+ H_1 & 10\%
\end{array}
$\quad \quad\quad& \quad\quad\quad
$
\begin{array}{clc} {\tilde t}_1 & \rightarrow t +LSP& 100\% \\
N_1 & \rightarrow t + {\tilde t}^* & 50\% \\
N_1 & \rightarrow {\bar t} + {\tilde t} & 50\% \\
{\tilde b}_1 & \rightarrow {\tilde t_1} +W^- & 100\% \\
{\tilde t}_2 & \rightarrow {\tilde t_1} +Z& 78\% \\
{\tilde t}_2 & \rightarrow {\tilde b_1} +W^+& 14\% \\
{\tilde t}_2 & \rightarrow {\tilde t_1}+ H_1 & 8\%
\end{array}
$
\end{tabular}
\end{center}
\caption{Branching fractions for benchmark spectra 1 and 2 with a ${\tilde t}$ NLSP.\label{decay12}}
\end{table}

\subsection{Spectum 1: stealth stop NLSP}

The first spectrum corresponds to a stealthy $\tilde t$
scenario \cite{Fan}, with the $\tilde{t}_1$ almost degenerate with the $t$.
The largest LHC SUSY production process in this scenario is
$pp\rightarrow {\tilde t}_1 {\tilde t}_1^*$ production, which is about
18 \% of the $t{\bar t}$ production cross section \cite{Beenakker}
at the 7 TeV LHC.  For a small enough gravitino mass, the ${\tilde t}_1$ decays
promptly with very little  missing transverse energy.  The ${\tilde t}_1$ can only
be uncovered by a precise measurement of the the $t{\bar t}$ cross section or a
shape analysis of the invariant mass distribution of  $t{\bar t}$ pairs.

The next largest SUSY production process are $pp\rightarrow {\tilde b}_1
{\tilde b}_1^*$ and $pp\rightarrow {\tilde t}_2 {\tilde t}_2^*$ which
are about few~$\cdot\, 10$ fb,  of the order of 0.1 \% of the $t{\bar t}$
production cross section \cite{Beenakker} at the 7 TeV LHC.  The experimental bounds on $\tilde{t}_2,\tilde{b}_1$ are around 270 GeV if decaying to $N_1,C_1$, while the bound from decays to light gravitinos/binos can be as high as 350 GeV~\cite{lightstopAndi}. The
${\tilde b_1}$ decays  to $\tilde t_1 W$, giving rise to $t {\overline
t} WW$ final states and in principle, missing energy.  The $N_1$ decays to $t {\tilde t_1}^*$, and the off-shell ${\tilde t}$'s will further decay to off-shell $t$'s. The final state for a pair production of $\tilde{t}_2$ will then contain $ttbb$ plus the decay products of two off-shell W's.
 The $\tilde{b}$  decays would be the best channels for looking for this spectrum. However all of these events will have very
little missing transverse energy. In the rest frame the gravitino will carry only a little energy.  Even though the lightest ${\tilde t}$ will be
boosted,  boost factors of order a few will generically not bring the
missing energy above the standard cuts.  As seen in Table \ref{decay12}, almost all superpartner decay chains end in the
NLSP, the ${\tilde t_1}$, which decays to $t$ and a soft gravitino. The
${\tilde t}$ lifetime is \cite{Sarid}
\beq
\Gamma = \frac{m_{\tilde t}^5}{16 \pi
F^2 }\left(1- \frac{m_t^2}{m_{\tilde t}^2}\right)^4~.
\eeq
For
$m_{\tilde t}< 200$ GeV, a prompt decay requires $\sqrt{F} $ less than
50 TeV, in which case there is no easy way to find a SUSY signal \cite{Kats} from this mode. For bigger values of $F$ there will be displaced vertices involving $t$
quarks.

\subsection{Spectrum 2: stop NLSP with heavier $N_1$}

The phenomenology with the second set of input parameters  is fairly
similar with a slightly heavier ${\tilde t}$. The main difference is that  we no longer have a stealth spectrum, $N_1$ is quite a bit heavier, and  more of the spectrum is pushed above a TeV.  Due to the heavier $N_1$ mass, it can now decay on shell to $t+\tilde{t}$, giving rise to events with $t {\overline t} t
{\overline t}$ in addition to the $t {\overline t} WW$ states from the $\tilde{b}$ decays.  Since the $\tilde{t}_1$ is still quite light, the amount of missing energy in these decays will still be limited. The $\tilde{t}_2$ will mainly decay to $Z+\tilde{t}_1$ giving rise to $t{\overline t}ZZ$ final states.

\subsection{Spectrum 3: minimal gauge mediation}

The third and fourth spectra both have neutralino (N)LSP's, thus the traditional missing energy signals of supersymmetry are expected.  However,  due to the heavy gluino and first two generations squarks, the rates are strongly reduced from those of the constrained MSSM. These spectra fall in the class of models considered in~\cite{lightstopAndi}.

The third set of input parameters in particular represent a minimal gauge mediated spectrum to the electric degrees of freedom.
All the soft scalar masses are set equal to 350 GeV.  Thus fixing
$m_{H_u}^2=m_{H_d}^2=(350 \,\,{\rm GeV})^2$ means that $f$ is no longer
really an input parameter but is an output of fixing the right EWSB vacuum.  Since we are considering gauge mediation, the expectation is that the LSP is again the gravitino, and the NLSP $N_1$ decays to photon plus gravitino. The lightest ${\tilde t}$ decays to $t^*N_1$, while the heavier ${\tilde t}$ has again many possible decay channels including $\tilde{t}_1Z, \tilde{b}W, N_{1,2,3}t, C_{1,2}b$, while the sbottom again decays to $\tilde{t}W$. Depending on the $N_1$ lifetime, the final states will again either be $j+$MET, $jt+$MET, and $j+W/Z+$MET, or the same final states with additional photons. This spectrum will also produce some longer SUSY cascades involving the same final states.

\subsection{Spectrum 4: high duality scale}

The fourth spectrum was chosen such that it can correspond to a higher duality scale, where the squark masses are mainly radiatively induced from the elementary gluino (and not coming from power suppressed terms), while the other composite soft masses  are small. In this case Higgs naturalness is especially good, since the Higgs soft breaking terms needed are around $(50\ {\rm GeV})^2$. Third generation squarks are in the 300-500 GeV range. The lightest ${\tilde t}$ decays via $\tilde{t}_{1}\to N_1c$, while the second ${\tilde t}$ has many possible decay modes to final states $\tilde{t}_1Z,C_1^+b, \tilde{b}W$ and $N_{1,2}t$. The sbottom decay is  $\tilde{b}_1\to \tilde{t}_1W$.  The characteristic final states will be  $j+$MET, $jt+$MET, or $jW/Z+$MET  events.   This yields fairly traditional SUSY signals at  reduced rate and no leptons (except from $W$ and $Z$'s).

\begin{table}
\begin{center}
\begin{tabular}{cc}
$
\begin{array}{|c|c|c|c|} \hline
N_1 & 88   \,\,\, {\rm GeV}   & C_2 & 415 \,\,\, {\rm GeV}  \\
H_1 & 128  \,\,\, {\rm GeV}  &  N_4 & 434 \,\,\, {\rm GeV}  \\
\tilde t_1 & 191  \,\,\, {\rm GeV}  &  H_2 & 473  \,\,\, {\rm GeV}  \\
N_2 & 192  \,\,\, {\rm GeV}  & \tilde t_2& 517  \,\,\, {\rm GeV}  \\
N_3 & 291  \,\,\, {\rm GeV}  & N_5 & 613  \,\,\, {\rm GeV}  \\
C_1 & 327  \,\,\, {\rm GeV}  & H^\pm & 650  \,\,\, {\rm GeV}  \\
{\tilde b}_1 & 350\,\,\, {\rm GeV}  & H_3 & 657  \,\,\, {\rm GeV}  \\
A_1 & 412 \,\,\, {\rm GeV}  & A_2 & 702 \,\,\, {\rm GeV}  \\
\hline
\end{array}
$\quad \quad\quad& \quad\quad\quad
$
\begin{array}{|c|c|c|c|}
\hline
H_1 & 126 \,\,\, {\rm GeV}  & N_2 & 348 \,\,\, {\rm GeV}  \\
A_1 & 190  \,\,\, {\rm GeV}  &  H_3 & 353 \,\,\, {\rm GeV}  \\
N_1 & 217 \,\,\, {\rm GeV}  & \tilde b_1 & 400  \,\,\, {\rm GeV}  \\
{\tilde t}_1 & 284  \,\,\, {\rm GeV}  &  A_2 & 460  \,\,\, {\rm GeV}  \\
H_2 & 339 \,\,\, {\rm GeV}  &   \tilde t_2 & 546 \,\,\, {\rm GeV}  \\
H^\pm & 341\,\,\, {\rm GeV}  & N_3 &  559   \,\,\, {\rm GeV}  \\
C_1 & 341\,\,\, {\rm GeV}  & N_4 & 602   \,\,\, {\rm GeV}  \\
\hline
\end{array}
$
\end{tabular}
\end{center}
\caption{Benchmark spectra 3 and 4.\label{masses34}}
\end{table}

\begin{table}
\begin{center}
\begin{tabular}{cc}
$
\begin{array}{clc}
{\tilde t}_1 & \rightarrow N_1^+ +b +W^+ & 100\% \\
{\tilde b}_1 & \rightarrow N_3 + b & 80\% \\
{\tilde b}_1 & \rightarrow {\tilde t}_1 + W^- & 95\% \\
{\tilde b}_1 & \rightarrow N_3 + b & 4\% \\
{\tilde b}_1 & \rightarrow N_1 + b & 1\% \\
{\tilde t}_2 & \rightarrow {\tilde t}_1 +Z & 42\% \\
{\tilde t}_2 & \rightarrow {\tilde b}_1+ W^+& 31\% \\
{\tilde t}_2 & \rightarrow N_2+ t & 10\% \\
{\tilde t}_2 & \rightarrow C_2^+ + b & 8\% \\
{\tilde t}_2 & \rightarrow N_1+ t & 4\%    \\
{\tilde t}_2 & \rightarrow C_1^+ + b & 3\% \\
{\tilde t}_2 & \rightarrow N_3+ t & 2\%
\end{array}
$\quad \quad\quad& \quad\quad\quad
$
\begin{array}{clc}
{\tilde t}_1 & \rightarrow N_1+c & 99\% \\
{\tilde t}_1 & \rightarrow N_1 +u & 1\% \\
{\tilde b}_1 & \rightarrow {\tilde t}_1 + W^- & 100\% \\
{\tilde t}_2 & \rightarrow {\tilde t}_1 +Z & 28\% \\
{\tilde t}_2 & \rightarrow C_1^+ + b & 24\% \\
{\tilde t}_2 & \rightarrow {\tilde b}_1+ W^+& 20\% \\
{\tilde t}_2 & \rightarrow N_2+ t & 15\% \\
{\tilde t}_2 & \rightarrow N_2+ t & 14\%
\end{array}
$
\end{tabular}
\end{center}
\caption{Branching fractions for benchmark spectra 3 and 4.\label{decay34}}
\end{table}

\begin{table}
\begin{center}
$
\begin{array}{c|c|c|c|c}
{\rm
SM\ fields}&	 {\rm spectrum}\,\,\, 1    & {\rm spectrum}\,\,\, 2 	& {\rm spectrum}\,\,\, 3		& {\rm spectrum}\,\,\, 4	\\ \hline
\gamma\gamma		&1.02					&	 1.02                         & 0.95 				& 0.85\\
{\rm gluons}		& 0.65				& 0.83                          &      0.82 				& 0.73\\
WW, ZZ			& 0.89				&  0.96				& 0.89					& 0.74 \\
u\bar{u}			&  0.72	            			& 1.0               		&   0.89                  			& 0.72 \\
d\bar{d}			&  1.01					& 0.91                     &     0.89					&  0.77
\end{array}
$
\caption{Ratio of Higgs couplings to  SM Higgs couplings for the same mass for the four benchmark spectra to various SM fields. \label{Higgscouplings}}
\end{center}
\end{table}

\section{Conclusions}
\label{sec:conclusions}
\setcounter{equation}{0}
\setcounter{footnote}{0}

We have seen that by combining supersymmetry, which makes the theory
calculable but also the Higgs too light and/or fine-tuned, with
compositeness, which requires strong coupling and allows for a heavier
Higgs with large dynamical Yukawa  couplings to other composites, we can
address three hierarchies: the hierarchy in Yukawa couplings, the little
hierarchy problem, and the apparent  hierarchy in squark soft masses. The strong
dynamics determines which particles have significant coupling to the
composite Higgs and can force the composite superpartners that are thus
required for naturalness to be much lighter than the elementary
superpartners.

In the model presented here Seiberg duality provides the crucial ingredient for resolving these hierarchies. The lessons could apply more generally but with Seiberg duality, we can explicitly determine the hierarchies in the spectrum of composite superpartners. The models we presented produce a composite Higgs, $t$ and LH $b$ along with partially composite $W$ and $Z$. The low energy dynamics is that of the NMSSM with a composite singlet, where the singlet couplings equal the $t$ Yukawa coupling. This ensures that the Higgs can be sufficiently heavy. The flavor problem is addressed via the large dynamical top Yukawa, and the little hierarchy via the NMSSM-type singlet coupling that determines the effective $\mu$-parameter and is related to the top Yukawa. The strong dynamics at the edge or just inside the conformal window will strongly suppress the soft breaking terms for the composites. This gives the necessary hierarchy among the squark masses, that will strongly reduce the SUSY production rates at the LHC and allow for a natural SUSY EWSB sector.

We have presented four distinct mass spectra corresponding to explicit implementations of this model. Two of them have the $\tilde{t}$ as the  NLSP (with gravitino LSP's), while the other two have the $N_1$ as the (N)LSP. One of the spectra with a $\tilde{t}$ NLSP correspond to an explicit implementation of a stealthy stop, where most of the SUSY events would not contain much missing energy.

Although conventional supersymmetric models are being challenged by experiments and naturalness at this point, this model raises the hope that models with more subtle composite dynamics could in fact be the correct theory of nature.

\section*{Acknowledgements}

We thank Markus Luty, Maxim Perelstein, Matt Reece, Yael Shadmi, Jessie Shelton, and Jay Wacker for useful discussions and comments.  We thank the KITP at UC Santa Barbara, where this work was initiated, for hosting us. The research of C.C. was supported
in part by the NSF grant PHY-0757868. L.R. was supported in part by the the NSF grant PHY-0855591. J.T. is supported by the US
Department of Energy  grant DE-FG02-91ER40674.

\end{document}